\documentstyle[aps,prl,multicol,epsf]{revtex}

\begin{document}
\def\CC{{\rm\kern.24em \vrule width.04em height1.46ex depth-.07ex
\kern-.30em C}}
\def\P{{\rm I\kern-.25em P}}
\def\RR{{\rm
         \vrule width.04em height1.58ex depth-.0ex
         \kern-.04em R}}

\draft
\title{  Stabilization
 of Quantum Information:
 A Unified Dynamical-Algebraic
 Approach
}
\author{Paolo Zanardi }

\address{
 Institute for Scientific Interchange  (ISI) Foundation,
Viale Settimio Severo 65, I-10133 Torino, Italy\\
Istituto Nazionale per la Fisica della Materia (INFM)
}
\date{\today}
\maketitle

\begin{abstract}
The notion of symmetry is shown to be at the heart
of all  error correction/avoidance
strategies  for preserving quantum coherence
of an open quantum system S
e.g., a quantum computer.
The existence of a non-trivial group of symmetries
of the dynamical algebra of S
provides state-space 
sectors immune to decoherence. 
Such noiseless sectors, that  can be viewed as a noncommutative version 
of the pointer basis,  are shown to support universal quantum computation 
and to be robust against perturbations.
When the required symmetry is not present one can generate  it artificially resorting 
to active symmetrization procedures. 
\end{abstract}

\section{Introduction}
Stabilizing  quantum-information   processing 
against
the environmental interactions as well as
operation imperfections
is a vital goal for
 any practical application of the protocols  of 
Quantum Information and Quantum Computation theory.
To the date  three kind of strategies  have been devised
in order to satisfy such a crucial requirement:
a) Error Correcting Codes  
which, in analogy with classical information theory, stabilize actively
quantum information by using redundant encoding
and measurements; b) Error Avoiding Codes  
pursue a passive stabilization
by exploiting symmetry properties  of the environment-induced noise
for suitable redundant encoding;
c)  Noise suppression schemes 
 in which, exploiting symmetry properties and  with no redundant encoding,
the unwanted  interactions are averaged away
by  frequently iterated external pulses.
We shall show how all these schemes
derive conceptually from  a unified  dynamical-algebraic framework.

Such an underlying  dynamical-algebraic structure  
is provided by the 
the reducibility of  the operator algebra
describing the faulty interactions 
of   the coding quantum system.
This property in turns  amounts to the existence of a non-trivial
group of symmetries for the global dynamics.
We  describe  a unified framework
which allows us to build systematically new classes of  error correcting
codes and noiseless subsystems.
Moreover we shall argue  how by using  symmetrization  strategies  one
can artificially produce noiseless subsystems 
and to perform universal quantum computation within 
these decoherence-free sectors.
\section{Noiseless Subsystems}
Suppose that $S$ is  quantum system coupled to an environment.
Without  further assumptions the decoherence induced by this coupling is likely to affect all the states of $S.$
Suppose now that $S$  turns out to be bi-partite, say $S=S_1+S_2,$ and moreover that
the environment is actually coupled just with $S_1.$
 If this is the case one can encode information is the quantum state of $S_2$
in a obvious noiseless way.

The  idea underlying the algebraic constructions  that will follow is conceptually  
nothing but an  extension of the extremely simple example above:
{\em the symmetry  of the system plus environment dynamical algebra provides
$S$ with an hidden multi-partite  structure  such that the environment is not able
to extract information out of some of these ``virtual'' subsystems.
}

Let $S$ be an open quantum system, with (finite-dimensional) state-space ${\cal H},$
and self-Hamiltonian $H_S,$ coupled to its environment  through the hamiltonian
$H_I= \sum_\alpha S_\alpha\otimes B_\alpha\neq 0,
$
where the $S_\alpha$'s  ($B_\alpha$'s) are system (environment) operators.
The unital associative algebra $\cal A$ 
closed under hermitian conjugation $S\mapsto S^\dagger,$
generated by the  $S_\alpha$'s
and $H_S$ will be referred to as the { interaction} algebra. 
In general $\cal A$ is a { reducible} $^\dagger$-closed 
subalgebra of the algebra $\mbox{End}({\cal H})$ of all the linear
operators over $\cal H:$
it   can be written as a 
direct sum of $d_J\times d_J$ (complex) matrix algebras each  one of which appears
with a multiplicity $n_J$
\begin{equation}
{\cal A}\cong\oplus_{J\in{\cal J}} {\bf{Id}}_{n_J}\otimes M(d_J,\,\CC).
\label{alg-split}
\end{equation}
where ${\cal J}$ is suitable finite set labelling the 
irreducible components of $\cal A.$
The  associated state-space decomposition  reads
\begin{equation}
{\cal H}\cong \oplus_{J\in{\cal J}} \CC^{n_J}\otimes \CC^{d_J}.
\label{split}
\end{equation}
These decompositions encode all  information about 
the possible quantum stabilization  strategies.

Knill et al  noticed that 
(\ref{alg-split}) implies that each factor $\CC^{n_J}$  in eq. (\ref{split})
corresponds to a sort of effective subsystem of $S$  coupled to the  environment
in a state independent way.
Such subsystems are then referred to as { noiseless}.
In particular one gets a { noiseless code} i.e., a decoherence-free subspace, 
${\cal C}\subset {\cal H}$ when in equation (\ref{split}) 
there appear one-dimensional irreps $J_0$ with multiplicity greater than one 
${\cal C}\cong \CC^{n_{J_0}}\otimes \CC$ .

The commutant of $\cal A$  is defined as 
${\cal A}^\prime$ in End$({\cal H})$ of $\cal A$
by  
${\cal A}^\prime:=\{ X\,|\, [X,\,{\cal A}]=0\}.$

{\em The existence of  a NS is equivalent to} 
\begin{equation}
{\cal A}^\prime
\cong
\oplus_{J\in{\cal J}}  M(n_J,\,\CC)\otimes {\bf{Id}}_{d_J}
\neq \CC\,{\bf{Id}}
\end{equation}
The condition ${\cal A}^\prime \neq \CC{\bf{Id}}$
is amounts to   the existence of a non-trivial
group of symmetries ${\cal G}\subset U\,{\cal A}^\prime.$
One has that the more symmetric a dynamics, the more
likely it   supports NSs.

When  $\{S_\alpha\}$ is a commuting set of hermitian operators.
 $\cal A$ is an abelian algebra and Eq. (\ref{split}) [with $d_J=1$]
is the decomposition of the state-space according the joint eigenspaces  
of the $S_\alpha$'s. 
The pointer basis  discussed in relation to the so-called
environment-induced superselection  is  nothing but an orthonormal 
basis associated to the resolution (\ref{split}).
Thus   {\em the  NS's  provide in a sense 
the natural noncommutative  extension  of the pointer basis.}

Now we discuss the relation between NSs and error correction.
The interaction algebra $\cal A$ has to to be thought of generated by error operators
and it is assumed to satisfy Eq. \ref{alg-split}.
Let $|J\lambda\mu\rangle \, (J\in{\cal J},\,\lambda=1,\ldots,n_J;\,\mu=1,\ldots,d_J)$
be an orthonormal basis associated to the decomposition (\ref{alg-split}).
Let
 ${\cal H}^J_\mu:=\mbox{span}\{|J\lambda\mu\rangle\,|\,\lambda=1,\ldots,n_J\},$
and let ${\cal H}^J_\lambda$ be defined analogously.
The next proposition shows that to each NS there is associated
a family of ECCs called  $\cal A$-{\em codes}. It is a simple consequence 
of (\ref{alg-split}) and of the definition of ECCs.

{\em
The   ${\cal H}^J_{\mu}$'s (${\cal H}^J_{\mu}$)  are   $\cal A$-codes ( ${\cal A}^\prime$-codes)
for any subset $E$ of  error operator such
that $\forall e_i, e_j\in E\Rightarrow e_i^\dagger e_j\in{\cal A}$
}

The standard stabilizer  codes are recovered when one considers
a $N$-partite qubit system,
and an abelian subgroup $\cal G$ 
of the Pauli group ${\cal P}.$
Let us consider the  state-space decomposition (\ref{split}) associated to $\cal G.$
If $\cal G$ has $k<N$ generators 
then $|{\cal G}|=2^k,$ whereas from commutativity  it follows  $d_J=1$ and $|{\cal J}|=|{\cal G}|.$
Moreover one finds
$n_J=2^{N-k}:$ each of the $2^k$ joint eigenspaces of ${\cal G}$ (stabilizer code)
 encode $N-k$ logical qubits.
It follows that
\begin{equation}
{\cal H}=\oplus_{J=1}^{2^k} \CC^{2^{N-k}}\otimes
\CC\cong  \CC^{2^{N-k}}\otimes  \CC^{2^k}.
\label{stabilizer}
\end{equation}
The allowed  errors belong  to the algebra ${\cal A}={\bf{Id}}_{2^{N-k}}\otimes M(2^{k},\,\CC).$
In particular errors  $e_i,\,e_j\in{\cal A}\cap{\cal P}$ are such that
 $e_i^\dagger e_j$ either belong to $\cal G$ or { anticommute}
with (at least) one element $\cal G.$
In this  latter case one has a 
 a non-trivial action on the $ \CC^{2^k}$ factor.

Let us finally stress out  that  the NSs approach described so far
works even in the case in which 
the interaction operators $S_\alpha$
represent, rather than coupling with external degrees of freedom,
internal unwanted {\em interna}l interactions i.e., $H_S^\prime= H_S +\sum_\alpha S_\alpha.$
\section{Collective Decoherence.}
Collective decoherence arises when a multi-partite quantum system,
is coupled symmetrically with a common environment.
This is the paradigmatic case for the  emergenge of noiseless subspaces and 
NS's as well.
More specifically
one has
 a $N$-qubit system  ${\cal H}_N:= (\CC^2)^{\otimes\,N}$ and the relevant interaction algebra
 ${\cal A}_N$ coincides with the algebra of completely symmetric operators
over ${\cal H}_N.$
The commutant ${\cal A}_N^\prime$ is the group algebra $ \nu (\CC {\cal S}_N),$
where $\nu$ is   the natural representation of the symmetric group ${\cal S}_N$  over ${\cal H}_N:$
$\nu(\pi) \otimes_{j=1}^N |j\rangle = \otimes_{j=1}^N |\pi(j)\rangle,\,(\pi\in{\cal S}_N).$
Uing elementary $su(2)$ representation theory one finds:

{\em
${\cal A}_N$
supports NS with dimensions }
\begin{equation}
n_J=\frac{(2\,J+1)\,N!}{(N/2+J+1)!\,(N/2-J)!}
\end{equation}
{\em where $J$ runs from $0$ ($1/2$) for $N$ even (odd).}
If in the above  ${\cal A}_N$ is replaced  by  its commutant,
the above result holds with $n_J=2\,J+1.$
Of course  collective decoherence allows for
${\cal A}_N$-codes as well.

In order to illustrate the general ideas let us consider r
$N=3.$
One has $(\CC^2)^{\otimes\,3}\cong \CC\otimes \CC^4+ \CC^2\otimes\CC^2.$
The last term  can be  written as
$\mbox{span}\{|\psi_\beta^\alpha\rangle\}_{\alpha\beta=1}^2
$
where
\begin{eqnarray}
|\psi_1^1\rangle &=& 2^{-1/2}(|010\rangle-|100\rangle),\,
|\psi_2^1\rangle =2^{-1/2}(|011\rangle-|101\rangle)\nonumber\\
|\psi_1^2\rangle &=&{2}/{\sqrt{6}}\,[1/2 (|010\rangle+|100\rangle)-|001\rangle],\nonumber\\
|\psi_2^2\rangle &=&{2}/{\sqrt{6}}\,[|110\rangle-1/2(|011\rangle+|101\rangle)].
\end{eqnarray}
The vectors  $|\psi_\beta^1\rangle$ and $|\psi_\beta^2\rangle$
( $|\psi_1^\alpha\rangle$ and $|\psi_2^\alpha\rangle$) span a two-dimensional ${\cal A}_3$-code
(${\cal A}^\prime_3$-code).
Taking the trace with respect to the index $\alpha$ ($\beta$)
one gets the ${\cal A}_3^\prime$ (${\cal A}_3$) NS's.
Moreover the first term corresponds to a trivial four-dimensional
${\cal A}_3^\prime$  code.
\section{NS synthesis  by Symmetrization .}
A typical  situation in which NSs could arise is when
the interaction algebra is contained in some reducible group representation
 $\rho$
of a finite  order (or compact) group $ {\cal G}$
Suppose that   the irrep decomposition of $\cal H$ associated to r $\rho$ has the form of Eq. (\ref{split})
in which
the  ${\cal J}$ labels a set of  ${\cal G}$-irreps $\rho_J$ (dim $\rho_J=d_J$).
 $\rho$ by extends linearly to the group algebra
$\CC{\cal G}:= \oplus_{g\in{\cal G}} \CC |g\rangle$
giving rise to a decomposition as in Eq. \ref{alg-split}.
It follows

{\em
If ${\cal A}\subset \rho(\CC\,{\cal G})$
then the dynamics supports (at least) $|{\cal J}|$ NS's
with dimensions $\{n_J(\rho)\}_{J\in{\cal J}}.$
}

The non-trivial assumption in the above statement is the reducibility of $\rho$
in that any subalgebra of operators belongs to a group-algebra.
As already stressed this is equivalent to a symmetry assumption.
When this required symmetry is lacking one can {\em artificially} generate it
by resorting  the so-called bang-bang techniques. 
These are  { physical} procedures,
involving iterated external ultra-fast  ``pulses'' $\{\rho_g\}_{g\in{\cal G}}$, whereby
one can synthesize, from a   dynamics generated by the $S_\alpha$'s,  
to a dynamics generated by $\pi_\rho(S_\alpha)$'s where
\begin{equation}
\pi_\rho\colon X\rightarrow\pi_\rho(X):=\frac{1}{|{\cal G}|}\sum_{g\in{\cal G}}
\rho_g\,X\,\rho_g^\dagger\in\rho(\CC{\cal G})^\prime.
\label{proj}
\end{equation}
 projects  any operator $X$ over the commutant of 
the algebra  $\rho(\CC{\cal G})$  generated by the bang-bang operations.
If one will, preserving the system self-dynamics, to get rid of unwanted  interactions $\S_\alpha$  with the environmnet 
he has  to  look for a  group ${\cal G}\subset U({\cal H}),$
such that 
i) $H_S\in \CC{\cal G}^\prime,$ ii) the  $S_\alpha$'s
transform according to non-trivial   irreps under the (adjoint) action of $\cal G.$
Then it  can be shown that
  $\pi_{\cal G}(S_\alpha)=0:$
 the effective dynamics of $S$ is unitary.

To understand how  this strategies might be useful for artificial NSs synthesis  is sufficient 
to notice that  Prop. 2  holds even for the  commutant by replacing
 the $n_J$'s with the $d_J$'s.
 Since the  $\cal G$-symmetrization of an operator 
belongs to  $\rho(\CC{\cal G})^\prime,$
one immediately finds that:

 {\em 
  ${\cal G}$-symmetrization of $\cal A$ supports (at least) $|{\cal J}|$ NS's  
with dimensions $\{d_J(\rho)\}_{J\in{\cal J}}.$}

It is remarkable that NSs do not allow just for safe encoding of quantum information but even
for its manipulation.
Form the mathematical point of view this result stems once again quite easily 
from the basic Eq. \ref{alg-split}which shows that the elements of ${\cal A}^\prime$ have non-trivial  action
over the $\CC^{n_J}$ factors.
Therefore: 

{\em If an experimenter has at disposal unitaries in $U {\cal A}^\prime$ 
universal QC is realizable within the NSs.
When such gates are not available from the outset they can be  obtained
through a ${\cal G}$-symmetrization of a couple of generic hamiltonians,
where ${\cal G}:=U\,{\cal A}$.}
\section{NS: Robusteness}
  In this section we prove a robusteness result for NSs extending analogous
 ones obtained for decoherence-free subspaces. Let ${\cal H}_S$ (${\cal H}_B$) denote the system state (environment) state-space.
Here $S$ represents the NS and the environment includes both the  coupled factor in Eq (\ref{split}) and the external degrees of freedom.

The evolution of the subsystem $S$ is given by
${\cal E}_\varepsilon^t(\rho) := \mbox{tr}_B [ e^{t\,{\cal L}_\varepsilon}(\rho\otimes\sigma)],$
where $\rho\in{\cal S}({\cal H}_S),\,\sigma\in{\cal S}({\cal H}_B).$
and the Liouvillian operator is given by  
 ${\cal L}_\varepsilon
 := {\cal L}_0+\varepsilon\,{\cal L}_1,$ where  $e^{t\,{\cal L}_0}$ acts trivially over $S$
i.e., $e^{t\,{\cal L}_0}(\rho\otimes\sigma)=\rho\otimes\sigma^\prime_t.$ In particular
$\rho\otimes{\bf{I}}_B$ is a fixed point.

The fidelity is defined as
\begin{equation}
 F_\varepsilon(t) := \mbox{tr}_S [\rho\,{\cal E}_\varepsilon^t(\rho)]= 
<\rho\otimes{\bf{I}}_B,\, e^{t\,{\cal L}}(\rho\otimes\sigma)>=\sum_{n=0}^\infty \varepsilon^n f_n(t).
\end{equation}
One has  $e^{t\,{\cal L}}=e^{t\,{\cal L}_0}\,{\cal E}^\varepsilon_t\,
e^{-t\,{\cal L}_0},$ in which, by defining 
${\cal L}_\varepsilon(\tau) :=e^{-\tau\,{\cal L}_0}\,{\cal L}_1\,e^{\tau\,{\cal L}_0}$ 
\begin{eqnarray}
{\cal E}_\varepsilon^t &:=&
e^{-t\,{\cal L}_0}\,e^{t\,{\cal L}}\,e^{t\,{\cal L}_0}={\bf{T}}\exp\left ( 
\int_0^t {\cal L}_1 (\tau)d\tau\right )\nonumber \\
&=&\sum_{n=0}^\infty \varepsilon^n
\int_0^t d\tau_1\cdots\int_0^{\tau_{n-1}}d\tau_n
\,{\cal L}_1^{\tau_1}\cdots{\cal L}_1^{\tau_n}.
\end{eqnarray}

\begin{eqnarray}
F_\varepsilon(t)&=&<\rho\otimes{\bf{I}}_B,\,e^{t\,{\cal L}}(\rho\otimes\sigma)>=
<\rho\otimes{\bf{I}}_B,\,e^{t\,{\cal L}_0}
{\cal E}_\varepsilon^t(\rho\otimes\sigma^\prime_t)>\nonumber \\
&=&<e^{t\,{\cal L}_0^\dagger}(\rho\otimes{\bf{I}}_B),
\,{\cal E}_\varepsilon^t(\rho\otimes\sigma^\prime_t)>=<\rho\otimes{\bf{I}}_B,\,
{\cal E}_\varepsilon^t(\rho\otimes\sigma^\prime_t)>
\nonumber \\
&=&\sum_{n=0}^\infty \varepsilon^n
\int_0^t d\tau_1\cdots\int_0^{\tau_{n-1}}d\tau_n
<\rho\otimes{\bf{I}},\,
{\cal L}_1^{\tau_1}\cdots {\cal L}_1^{\tau_n} (\rho\otimes\sigma^\prime_t)>\nonumber \\
&=& 1+  \varepsilon \int_0^t d\tau <\rho\otimes{\bf{I}}_B,\,{\cal L}_1^\tau (\rho\otimes\sigma^\prime_t) >
+o(\varepsilon^2)\nonumber \\ &=&
1+ \varepsilon \int_0^t d\tau  <\rho\otimes{\bf{I}}_B,\,{\cal L}_1
(\rho\otimes\sigma^{\prime\prime}_\tau)>+
o(\varepsilon^2)=1+o(\varepsilon^2)\nonumber
\end{eqnarray}
Where, for obtaining the last equality, we assumed  
\begin{equation}
<\rho\otimes{\bf{I}}_B,\,{\cal L}_1 (\rho\otimes\sigma^{\prime\prime}_\tau)>=0.
\label{0}
\end{equation}
We assume that the infinitesimal generator ${\cal L}$ of the dynamical semi-group
has the following standard (Lindblad) form that holds for Markovian evolutions 
\begin{equation}
{\cal L}(\rho):=\frac{1}{2}\sum_\mu ( [L_\mu \rho,\,L_\mu^\dagger]+ [L_\mu,\,\rho L_\mu^\dagger]).
\label{lin}
\end{equation}
Perturbing the Lindblad operators $L_\mu\mapsto L_\mu + \varepsilon\,\delta L_\mu,$ 
one gets ${\cal L}\mapsto {\cal L} +\varepsilon\, [ {\cal L}_1 + \varepsilon\, {\cal L}_2],$
where 
\begin{eqnarray}
{\cal L}_1(\omega) := \frac{1}{2}\sum_\mu ( [\delta L_\mu \omega,\,L_\mu^\dagger]+ 
[\delta L_\mu,\,\omega L_\mu^\dagger]+ [ L_\mu \omega,\,\delta L_\mu^\dagger]+ 
[L_\mu,\,\omega  \delta L_\mu^\dagger]),
\label{lin1}
\end{eqnarray}
\begin{eqnarray}
{\cal L}_2(\omega) := \frac{1}{2}\sum_\mu ( [\delta L_\mu\omega,\,\delta L^\dagger_\mu] + 
[\delta L_\mu,\,\omega\delta L^\dagger_\mu]).
\label{lin2}
\end{eqnarray} 
Moreover
$L_\mu := {\bf{I}}_S\otimes B_\mu,$ and $ \delta L_\mu := X_\mu \otimes A_\mu.$
Let us consider the first two terms of Eq. (\ref{lin1}) for a given $\mu$
and with $\omega=\rho\otimes\sigma$
\begin{eqnarray}
 & & 2\,\delta L_\mu\, (\rho\otimes\sigma)\, L_\mu^\dagger- L_\mu^\dagger\delta L_\mu\,(\rho\otimes\sigma)
-(\rho\otimes\sigma)\, L_\mu^\dagger\delta L_\mu\nonumber \\
&=& 2\, X_\mu\rho\otimes A_\mu\sigma L_\mu^\dagger - X_\mu\rho\otimes L_\mu^\dagger A_\mu \sigma-
\rho X_\mu \otimes \sigma L_\mu^\dagger A_\mu 
\end{eqnarray}
multiplying by $\rho\otimes{\bf{I}}_B$  and 
taking the trace
\begin{eqnarray}
\mbox{tr}_S (\rho X_\mu \rho)\,\mbox{tr}_B ( 2\,A_\mu\sigma L_\mu^\dagger-L_\mu^\dagger A_\mu \sigma
-\sigma L_\mu^\dagger A_\mu)=0.
\end{eqnarray}
Reasoning in the very same way, even the last terms of Eq. (\ref{lin1}) 
give a vaninshing contribution. This show that relation (\ref{0}) is fulfilled
by ${\cal L}_1$.
\section{Conclusions.}
The possibility of noiseless enconding and processing of quantum information
is traced back to the existence of an underlying multi-partite structure.
The origin 
of such  hidden  structure is purely algebraic
and it is dictated by the interactions between the systems and the environmentr:
When the latter  admits  non trivial symmetry group then {\em noiseless subsystems} allowing for 
 universal quantum computation 
exist. This  NSs approach, introduced by Knill et al 
as a generalization of decoherence-free subspaces, is robust against perturbations
and  provides an analog of the pointer basis in the  noncommutative realm.
The notion of NS has been shown to be crucial    for a unified understanding and designing 
of  e error correction/avoidance strategies.
In particular  we argued how one can use decoupling/symmetrization techniques 
for artificial synthesis  of systems  supporting NSs, and how to perform
on such NSs non-trivial computations.

\end{document}